\documentclass[12pt]{article}
\usepackage{latexsym}
\usepackage{graphicx}
\usepackage{amsmath}
\usepackage{hyperref}
\usepackage{amssymb}

\begin{document}

\begin{center}
\textbf{\Large The class of second order quasilinear  equations: models, solutions and background of classification } \vspace{0.5 cm}

 O. Makarenko$ ^{\dagger,}$\footnote{e-mail:
\url{makalex51@gmail.com}}, A. Popov$ ^{\ddagger,}$\footnote{e-mail:
\url{popovanton567@gmail.com}}, S. Skurativskyi$^{\S,}$\footnote{e-mail: \url{
skurserg@gmail.com}} \vspace{0.5 cm}

$ ^{\dagger}$Institute for Applied System Analysis NTU ''KPI''

Politekhnichna st., 14, 14B,  Kyiv, Ukraine, 03056

$ ^{\ddagger}$Institute of Physics and Technology NTU ''KPI''

Prosp. Peremohy, 37, Kyiv, Ukraine, 03056

$ ^{\S}$Subbotin institute of geophysics, Nat. Acad. of Sci. of Ukraine

     Bohdan Khmelnytskyi str. 63-G, Kyiv, Ukraine
\end{center}

\begin{quote} \textbf{Abstract.}{\small 
The paper is concerned with the unsteady solutions to the model of mutually penetrating continua and quasilinear hyperbolic modification of the Burgers equation (QHMB). The studies were focused on the peculiar solutions of models in question. 
On the base of  these models and their solutions, the ideas of second order quasilinear models classification were developed.  
}
\end{quote}

\begin{quote} \textbf{Keyword: }{\small hyperbolic equations, attractors, multivaluedness
 }
\end{quote}

\vspace{0.5 cm}

\section*{Introduction}

Nonlinear models for phenomena and systems are the cornerstones in modern physics. The examples of these models are well known, namely the Burgers, KdV, Liouville, nonlinear Shredinger  equations and etc. The derivation of analytical solutions for these equations is a challenge for scientists. Therefore, the numerical treatments of such models   are developed intensively. 

Among the models mentioned above it is worth  accentuating the second order in time nonlinear hyperbolic differential equations \cite{monog,Makarenko_2012} having a broad applications recently. Note that the partial solutions for these equations had been derived \cite{Popov_Mak_2015}.

Due to the significance of considered equation and their solutions, this paper deals with the classification of these equations, definition of general expression of quasilinear models. Models' solutions obtained via the numerical modelling are analyzed in detail. We also discuss the possible ways of investigations, in particular,  combining the concept of dynamical systems (attractors) and artificial intelligence methods (neural networks). The aspects related to the multivaluedness of solutions including symmetries are considered as well.

	\section{Wave regimes in media with oscillating inclusions}
	
To begin with, let us note that the class of second order quasilinear models is not empty  and covers  many different  models originated from the physics and biology. In particular, consider the model of mutually penetrating continua which uses for the description of physical processes in complex media \cite{nasu,nonlin_dym,Dan_Skur_2016}. It turned out that this model possesses the specific wave solutions. Consider these solutions in more detail. 

The model we are going to deal with has the following form  
\begin{equation}\label{model}
\rho \frac{{\partial ^2 u}}{{\partial t^2 }} = \frac{{\partial
\sigma }}{{\partial x}} - m \rho  \frac{{\partial ^2 w
}}{{\partial t^2 }}, \qquad \frac{{\partial ^2 w }}{{\partial t^2
}} + \Phi\left( {w - u} \right) = 0,
\end{equation}
where $\rho$ is  medium's density, $u$ and $w$ are the displacements of carrying medium and oscillator from the rest state,  $m\rho$ is the density of oscillating continuum.  We also use  the cubic constitutive equation for the carrying medium
 $
 \sigma  = e_1 u_x  + e_3 u_x^3
$, where $e_1$, $e_3$ are the  elastic moduli, and relation for applied force $ \Phi(x)=\omega ^2 x+\delta x^3$, where $\omega$  denotes the natural frequency of oscillator, whereas  the parameter $\delta$ appears  due to accounting for the cubic term in the expansion of restoring force in a power series.

The  traveling wave solutions of model (\ref{model}) have the following form
\begin{equation}\label{solution}
u=U(s), \quad w=W(s), \quad s=x-Dt,
\end{equation}
 where the parameter $D$
 is a  constant velocity of the wave front. 

 Inserting (\ref{solution}) into  model (\ref{model}), it is easy to see that  the   functions $ U$ and $ W$
 satisfy the dynamical system 
\[
D^2 U' = \rho^{-1}\sigma \left( {U'} \right) - mD^2 W^\prime,
\quad W'' + \Omega ^2 \left( {W - U} \right) +\delta D^{-2}\left(W-U\right)^3= 0,
\]
where $\Omega  = \omega D^{-1}$.

This system can be written in the form 
\begin{equation}\label{syst3}
\begin{split}
W' =\alpha _1 R + \alpha _3 R^3, \qquad
  U' =R, \qquad (\alpha_1+3\alpha_3 R^2)R'+\Omega^2(W-U)+\delta D^{-2}(W-U)^3=0,
\end{split}
\end{equation}
 where $\displaystyle \alpha _1  =
\frac{{e_1 - D^2 \rho }}{{m \rho D^2 }}$, $\displaystyle \alpha _3
= \frac{{e_3 }}{{m \rho D^2 }}>0$. Through the report we fix $$e_1=\rho=1, \,e_3=0.5, \,m=0.6, \,\omega=0.9$$ in numerical treatments.

At first, consider system (\ref{syst3}) at $\delta=0$. 
The detail description  of  phase plane of dynamical system had been done in the paper \cite{skur_akust}, we thus summarize 
the main results only.

\begin{figure}
\begin{center}
\includegraphics[width=7.5 cm, height=2.0 in]{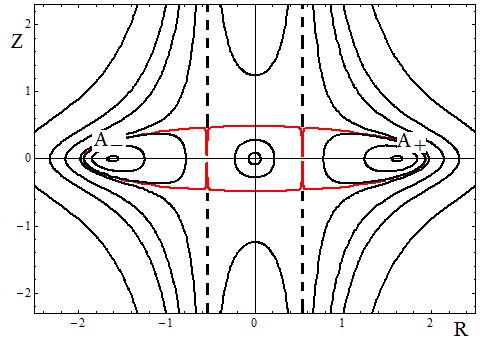} \hspace{0.4 cm}
\includegraphics[width=7.5 cm, height=2.0 in]{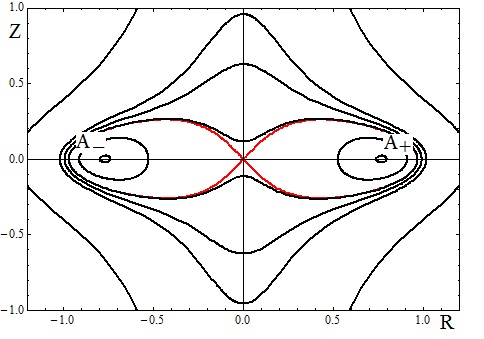}\\
(a)  \hspace{7 cm} (b)
\caption{The phase portraits for  system (\ref{syst3}) at (a): $\alpha_1<0$ ($D=1.2$) and (b): $0<\alpha_1<1$ ($D=0.9$).}\label{fig:1}
\end{center}
\end{figure}

At $\alpha_1<1$ the phase plane contains three fixed points, whereas at $\alpha>1$  the only one fixed point (center) remains. 
For  $\alpha_1<0$, when  $D=1.2$ is fixed for definiteness,  a typical phase portrait is depicted in the Fig.~\ref{fig:1}a. In this case, all fixed points are centers surrounded
by periodic orbits. There are  separatrices separated the regions with periodic and unbounded trajectories and two lines corresponding to discontinuity of system.

When $0<\alpha<1$, at $D=0.9$ for instance, in the phase portrait (Fig.~\ref{fig:1}b) one can distinguish   the homoclinic trajectories that go through 
the origin. The homoclinic loop can be written in the explicit form 
\begin{equation}\label{soliton}
\begin{array}{c}
\displaystyle  s - s_0  = \frac{{3 }}{2\Omega}\arcsin \left(
{\frac{{4\alpha _3 R^2 - 3 + 4\alpha _1 }}{{\sqrt {9 - 8\alpha _1 }
}}} \right) - \frac{1}{2\Omega} \sqrt{
\frac{\alpha _1 }{1 - \alpha _1 }} \times\\ \\
\displaystyle \ln \left(
 \frac{1}{R^2} + \frac{{3\alpha _3  - 4\alpha _1 \alpha _3 }}{{4\left( {\alpha _1  - \alpha _1^2 } \right)}} +
 \sqrt {\left\{ {\frac{1}{R^2} + \frac{{3\alpha _3  - 4\alpha _1 \alpha _3 }}{{4\left( {\alpha _1  - \alpha _1^2 } \right)}}} 
\right\}^2  - \frac{{\alpha _3^2 \left( {9 - 8\alpha _1 } \right)}}{{16\left( {\alpha _1  - \alpha _1^2 } \right)^2 }}}
 \right)  \Biggr|_{R_0^2 }^{R^2}
\end{array}
\end{equation}

 Solution (\ref{soliton}) corresponds to the solitary wave solution
with infinite support.

When $\alpha_1$ tends to zero, the angles between separatrices of saddle point $O$ are growing. As a result,   at  $ \alpha _1 = 0 $  
we observe  the transformation of  solitary wave into the compacton, i.e.,  solutions with finite support \cite{Hyman-Rosenau, vsan_lodz2016}. 
These orbits are described by the following expressions
\begin{equation*}
  U^\prime _s=\begin{cases}
   \sqrt {\frac{3}{{2\alpha _3 }}} \sin \frac{{\Omega s}}{3}, & \mbox{if} \quad \frac{{\Omega s}}{3} \in \left[ {0;\pi } \right]  \\
    0, & \mbox{if} \quad \frac{{\Omega s}}{3} \notin \left[ {0;\pi } \right]  \\
\end{cases}
\quad      \mbox{  and  } \quad
U=\begin{cases}
0, & \mbox{if}  \quad \frac{{\Omega s}}{3} \in \left[ -\infty; 0  \right],\\
\frac{3}{\Omega}\sqrt{\frac{3}{2\alpha_3}}\left(1-\cos\frac{{\Omega s}}{3}\right), &  \mbox{if}  \quad \frac{{\Omega s}}{3} \in \bigl(0; \pi  \bigl],\\
\frac{6}{\Omega}\sqrt{\frac{3}{2\alpha_3}}, &  \mbox{if}  \quad \frac{{\Omega s}}{3} \in \bigl(\pi; \infty  \bigl).\\
\end{cases}
\end{equation*}

As above, there are  the periodic orbits enclosed in the homoclinic loops  and  periodic trajectories l lying  beyond the homoclinic contour.

\subsection{Phase diagrams in the model with cubic nonlinearity in the equation of motion for oscillating inclusions}\label{sec12}

If $\delta\not = 0$, then system (\ref{syst3}) does not reduce to the dynamical system in the plane $(R;R')$. But the first integral  for  (\ref{syst3})  can still be derived in the form
\begin{equation}\label{skur:hamilton}
I=\frac{\mu_1}{2}\left(W-U\right)^4 + \mu_2\left(W-U\right)^2+ f(R),
\end{equation}
where $\mu_1=\delta D^{-2}$, $\mu_2=\omega^2 D^{-2}$, $f(R) = \alpha _3^2 R^6  + \frac{{4\alpha
_1 - 3}}{2}\alpha _3 R^4  + \left( {\alpha _1^2  - \alpha _1 }
\right)R^2$. Since $dI/ds=0$ on the trajectories of system (\ref{syst3}), then $I \equiv \mbox{const}$.

It is easy to see that expression (\ref{skur:hamilton}) can be used for splitting  system  (\ref{syst3}). Indeed, solving (\ref{skur:hamilton})
with respect $W-U$ we obtain
\begin{equation}\label{skur:branch}
W-U=\pm\sqrt{\frac{-\mu_2 \pm \sqrt{\mu_2^2-2\mu_1 f(R)+2\mu_1 I}}{\mu_1}}.
\end{equation}
This allows us to separate the last equation of (\ref{syst3}) from other ones. Unfortunately, the resulting equation cannot be 
integrated for general case, therefore,  let us consider its phase plane structure, which is equivalent to the structure of  level curves for function (\ref{skur:branch}).

\begin{figure}[h]
\begin{center}
\includegraphics[width=7.5 cm, height=2.0 in]{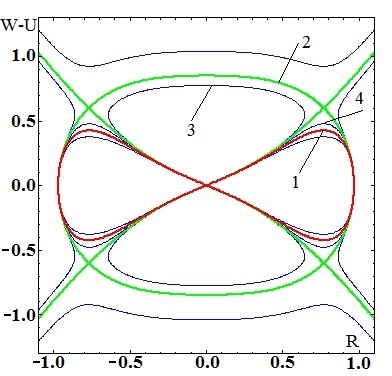}
\hspace{0.4 cm}
\includegraphics[width=7.5 cm, height=2.0 in]{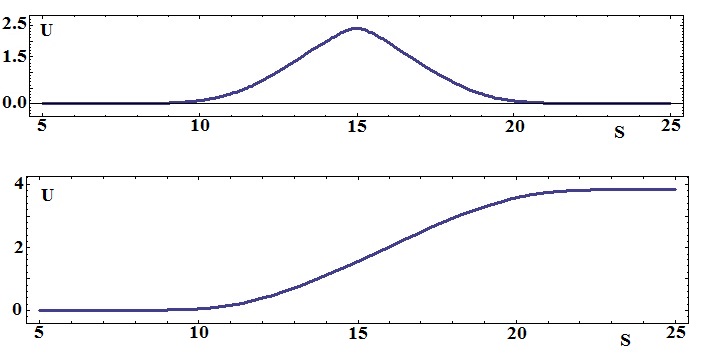}\\
(a) \hspace{8 cm} (b)
\caption{Left: Position of level curves  $I(\delta)=0$  at different values of $\delta$. Curve 1 is plotted at $\delta=0$, curve 2 
at $\delta_0$, curve 3  at  $\delta=-2.7<\delta_0$, curve 4  at $\delta=-1.5>\delta_0$.  Right: Homoclinic trajectories from the left diagram 
corresponding to $\delta=-2.7<\delta_0$ (upper panel) and $\delta=-1.5>\delta_0$ (lower panel). }\label{fig:2}
\end{center}
\end{figure}

Now consider the position and the form of homoclinic trajectories when the parameter $\delta$ is varied. 
Starting from the loop 
at $\delta=0$ which coincides with the orbits of Fig.~\ref{fig:1}b, we see that increasing $\delta$ causes the attenuation
 of  loop's size along vertical axis.  Actually, the level curve consists of the closed curve (homoclinic loops) and unbounded trajectories. 
If $\delta$ decreases,  loop's size grows, but at $\delta_0$ the additional heterocycle connecting four new saddle points appears.
 The bifurcational value $\delta_0$  can be derived via  analyzing the function (\ref{skur:branch}). Namely, $\delta_0$ corresponds to the moment when different branches of  (\ref{skur:branch})
are tangent. This happens when $\mu_2^2-2\mu_1 f(R)=0$. Thus, the condition of contact for two branches leads us to a cubic equation with respect to $R^2$  with zero discriminant. 
Then $\delta_0=-\frac{\alpha_3\omega^4 }{D^2(\alpha_1-1)^2}$ or $\delta_0=\frac{27 \alpha_3\omega^4 }{D^2\alpha_1^2 (9-8\alpha_1)}$. The last value of $\delta_0$
is not interesting because four branches of (\ref{skur:branch}) degenerate into two ones forming homoclinic loops.

Considering the first value $\delta_0$, we put $D=0.9$ and derive  $\delta_0=-2.24$. For $\delta>\delta_0$ we have homoclinic loops placed along horizontal axis accompanied by  
appearing  the unbounded curves  in the upper and lower parts of diagram. When $\delta<\delta_0$, the homoclinic loops are placed in the vertical quarters of the phase plane, whereas the unbounded orbits appear 
at the left and at the right sides of diagram (Fig.~\ref{fig:2}a).
Note that the profiles of the resulting solitary waves are different (Fig.~\ref{fig:2}b), namely, at $\delta<\delta_0$ the $U$ profile looks like a bell-shape curve, 
but at $\delta>\delta_0$ it is a kink-like regime.  

\begin{figure}[h]
\begin{center} 
\includegraphics[width=7.5 cm, height=5 cm]{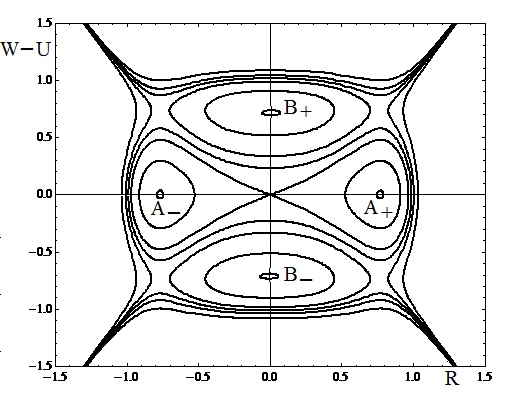}\hspace{0.5 cm}
\includegraphics[width=7.5 cm, height=5 cm]{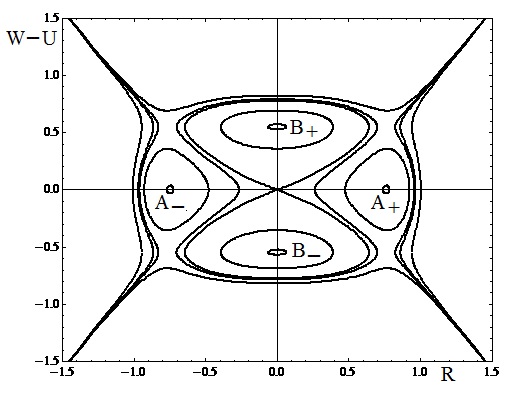}\\
(a) \hspace{8 cm} (b)
\caption{Phase portraits at $\delta=-1.5$(a) and $\delta=-2.7$(b). 
}\label{skur:phasdelta}
\end{center}
\end{figure}

The homoclinic orbits divide the phase plane into parts filled by closed curves corresponding to the periodic regimes.  If we choose  $\delta=-1.5>\delta_0$, we get  the typical phase 
portrait of system (\ref{syst3}) plotted in 
the Fig.~\ref{skur:phasdelta}a. In the portrait  two pairs of nontrivial fixed points  $A_\pm (\pm Q;0)$ and $B_\pm (0; \pm \omega/\sqrt{-\delta})$ can be distinguished. Inserting the coordinates of these points into 
relation (\ref{skur:hamilton}),  we obtain the values of $I_1=-\frac{(\alpha_1-1)^2}{2\alpha_3}$ and $I_2=\frac{\omega^4}{2D^2|\delta|}$ which allows us to state the conditions of periodic regimes existence. 
For fixed $\delta=-1.5$, $I_1=-0.18$ and $I_2=0.27$. We thus get that periodic regimes exist if $I_1<I<I_2$ only. 

Now let us choose  $\delta=-2.7<\delta_0$. In this case the homoclinic loop going through the origin is placed along vertical axis and  the phase portrait looks like  Fig.~\ref{skur:phasdelta}b. As above, we have $I_1=-0.18$, but $I_2=0.15$.
 
\subsection{Wave dynamics of   model (\ref{model})}

\noindent  To model the wave dynamics, we used  the three level
finite-difference numerical scheme for model (\ref{model}). 

\begin{figure}[h]
\begin{center} 
\includegraphics[width=10 cm, height=7 cm]{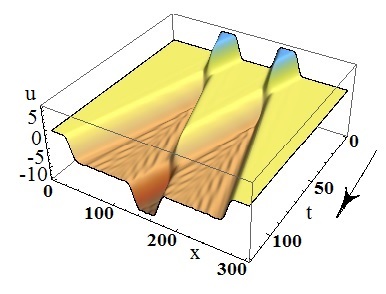}\\
\caption{ The propagation of  solitary waves at $\delta=0$ starting from the two-arch initial profile.}\label{skur:fig4}
\end{center}
\end{figure}

{\bf Solitary waves.} To consider the evolution of solitary waves, let us construct the initial data $v_i$, $q_i$, $G_i$, $F_i$ for numerical simulation on the base of homoclinic contour. 
To do this, we integrate dynamical system (\ref{syst3}) with initial data $R(0)=10^{-8}$, $Z(0)=0$, $s \in [0;L]$  and choose the right  homoclinic loop in the phase portrait (Fig.~\ref{fig:1}b). 
Then the profiles of $W(s)$, $U(s)$, and $R(s)$ can be derived. Joining the proper arrays, we can build the profile in the form of  arch:
$$
v=U(ih)\cup U(L-ih),\, q=u(x+\tau D)=\left[U(ih)+\tau D R(ih)\right]\cup \left[U(L-ih)+\tau D R(L-ih)\right].
$$

The arrays $G$ and $F$ are formed in similar manner. 
Combining two arches and continuing the steady solutions 
at the ends of graph, we get more complicated profile.
 We apply the fixed boundary conditions, i.e. $u(x=0,t)=v_1$,   $u(x=Kh,t)=v_K$, where $K$ is the length of an array.

Starting from the two-arch initial data, we see (Fig.~\ref{skur:fig4}) that solitary waves move to each other, vanish during approaching, and appear with negative
 amplitude and shift of phases. After collision in the zones between  waves some ripples are revealed.  Secondary collisions of waves are watched also. Note that the
 simulation of compacton solutions displays similar properties. 

\begin{figure}[h]
\begin{center} 
\includegraphics[width=7.5 cm, height=6.0 cm]{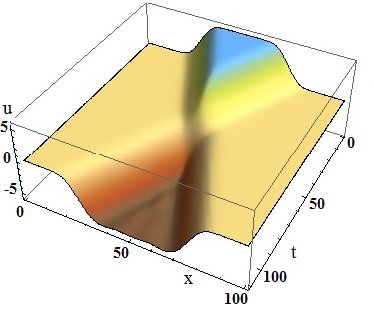}\hspace{0.4cm}
\includegraphics[width=7.5 cm, height=5.0 cm]{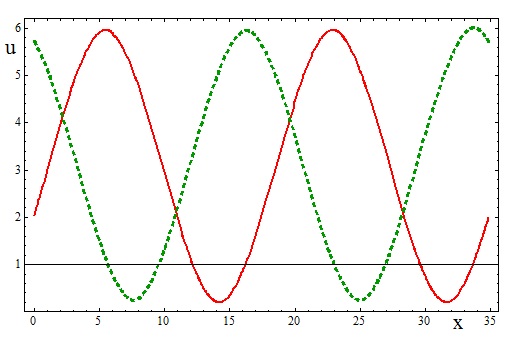}\\
(a) \hspace{8 cm} (b)
\caption{Collision of solitary waves (a) and evolution of periodic wave (b) at $\delta=-0.2$. }\label{skur:fig5}
\end{center}
\end{figure}

Propagation of solitary waves at $\delta \not =0$ depends on the sign of $\delta$. Numerical simulations  show that the 
collision of waves at  $\delta>0$ is similar to the collision at  $\delta=0$. Behavior of waves after interaction does not change essentially  when  $\delta<0$ and close to zero (Fig.~\ref{skur:fig5}a). 
But if  $\delta$ is not small,  after collision the amplitude of solution is increasing in the place of soliton's intersection  and  after a while   the solution  is  destroyed \cite{dan_lodz_16}. 

 This suggests that we encounter  the unstable  interaction of solitary waves or the numerical scheme we used possesses spurious solutions. But if we take 
half spatial step and increase the scheme  parameter $r$ up to 0.8, the scenario of solitary waves collision is not changed qualitatively. Therefore, the assertion on the unstable nature of collision is more preferable.

{\bf Periodic waves.} To simulate the evolution of periodic waves, we take the initial profile corresponding to a periodic orbit (red curve in Fig.~\ref{skur:fig5}b)
surrounding the homoclinic loop at the phase portrait of Fig.~\ref{skur:phasdelta}a   at $\delta=-0.2$ and $D=0.9$.  Due to the periodicity of the problem, 
the boundary conditions for numerical scheme  should be modified \cite{dan_lodz_16}.
The resulting profile derived after time interval $1200\tau$  has passed is depicted in Fig.~\ref{skur:fig5}b with dark green curve. It is obvious that this wave is shifted  
a distance $1200\tau D$ to the right in a self-similar manner.

\section{Unsteady solutions to QHMB}

	Let us consider the evolution of localized perturbation within the framework of 	QHMB:
	\begin{equation}\label{QHMB}
	\tau \frac{\partial ^2 u}{\partial t^2}+\alpha\frac{\partial u}{\partial t}+\beta\varphi(u)\frac{\partial u}{\partial x}=\mu k(u)\frac{\partial^2 u}{\partial x^2}+\nu \psi (u) \left(\frac{\partial u}{\partial x}\right)^2+\theta f(u),
	\end{equation}
	where $\tau$, $\alpha$, $\beta$, $\mu $, $\nu$, $\theta$ are constants.
	
\begin{figure}[h]
\begin{center} 
\includegraphics[width=7.5 cm, height=5.0 cm]{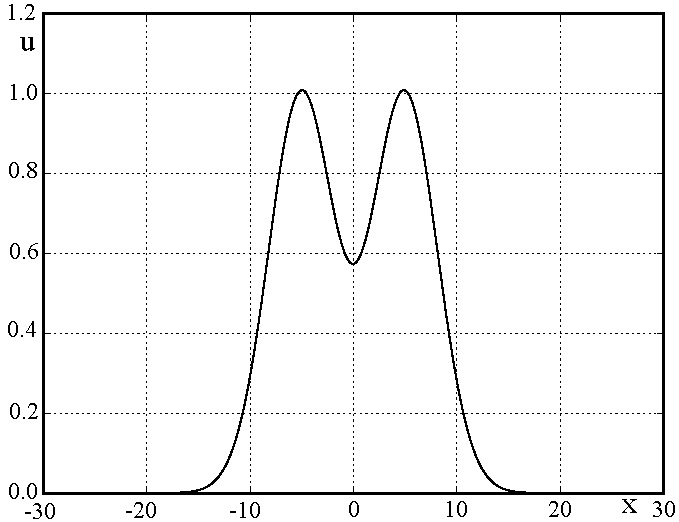}\hspace{0.4cm}
\includegraphics[width=7.5 cm, height=5.0 cm]{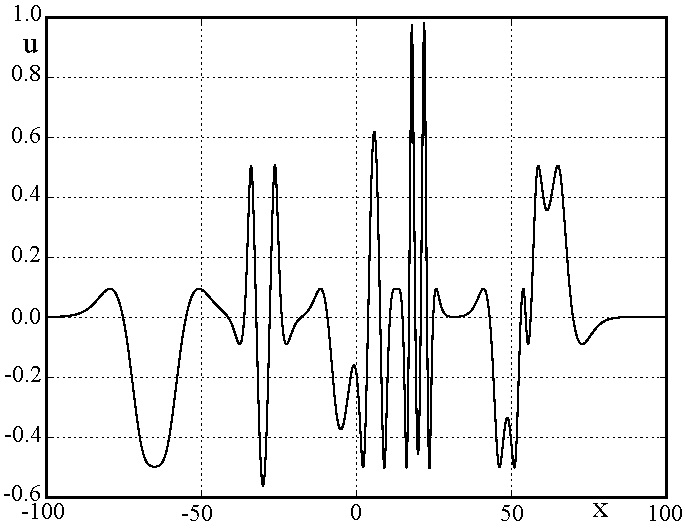}\\
(a) $t=0$ \hspace{6 cm} (b) $t=0.505$\\
\includegraphics[width=7.5 cm, height=5.0 cm]{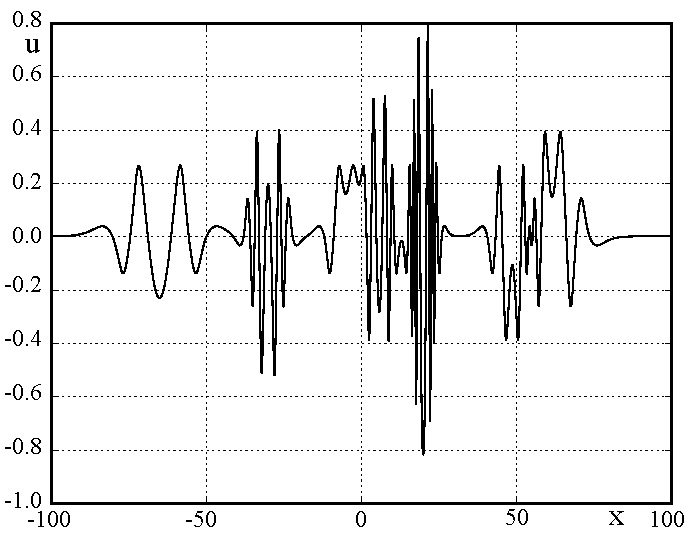}\hspace{0.4cm}
\includegraphics[width=7.5 cm, height=5.0 cm]{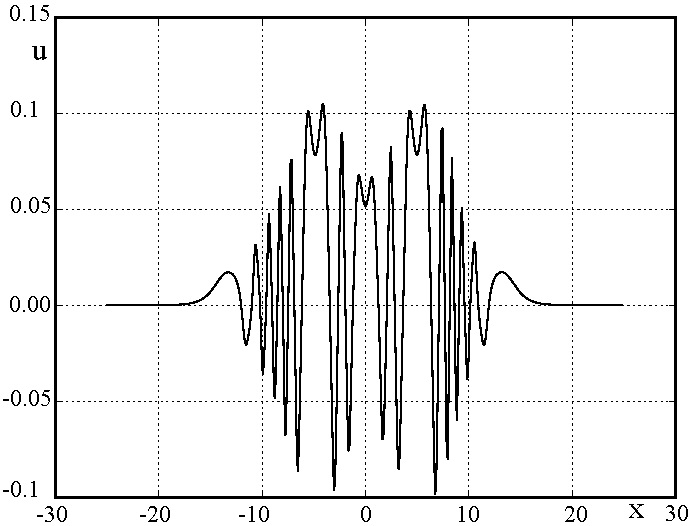}\\
(c) $t=1.56$ \hspace{6 cm} (d) $t=6.875$
\caption{The initial profile for model (\ref{QHMB}) and its evolution.}\label{smp:figQHMB}
\end{center}
\end{figure}

	As an initial data for equation (\ref{QHMB})б we chose the profile 
	$u(x,0)=\exp\left(\frac{x+a}{b}\right)+\exp\left(\frac{x-a}{b}\right)$ (fig.~\ref{smp:figQHMB}a), where $a=5$, $b=4.5$. The evolution of this two-hump perturbation is shown in figs.~\ref{smp:figQHMB}b,c,d.
	
From the analysis of these figures it follows that  the number of oscillations increases during evolution. The oscillations have non-regular character but remain localized in spatial domain. Thus, one can call them the ``pre-turbulent oscillations'' or, due to the localization, the ``turbulons''. Let us also note that the appearance of this type solutions was discovered in other numerical simulations of $2D$ and $3D$ problems \cite{monog}.
	
\section{The class of quasilinear equations and their formal description}\label{mps:sec_one}

The presented examples of modelling the carrying processes manifest the variety of solutions of models considered. Analyzing these models in details, one  can be convinced that they belong to the second order quasilinear equations, i.e., the relation is a linear with respect to higher derivatives whereas it is nonlinear with respect to lower derivatives, unknown functions,  and independent variables.    It is obvious that this class of equations is very broad and needs to systematize.

Let us start from the defining the general expression of quasilinear second order models, namely 
\begin{equation}\label{mps:system1}
 \begin{split}
\tau r(u) \frac{\partial ^2 u}{\partial t^2}&+\alpha s(u) \frac{\partial  u}{\partial t}+\beta \varphi (u) \frac{\partial u}{\partial x} = \mu k(u)  \frac{\partial ^2 u}{\partial x^2} +\nu \psi (u) \left( \frac{\partial  u}{\partial x}\right)^2 + \gamma h(u)  \frac{\partial ^2 u}{\partial x\partial t} +\\
&+\xi b(u) \left( \frac{\partial  u}{\partial t}\right)^2+\theta f(u)+ \chi I(u,x,t,\pm \Delta t, \pm\Delta x;...),
 \end{split}
\end{equation}
where $I$ is the source (integral form addmited), 
$\tau$, $\alpha$, $\beta$, $\mu $, $\nu$, $\theta$ are constants, $r$, $s$, $k$, $\psi$, $h$, $b$, $f$ are the specified functions.

Note that, depending on the coefficients and functions, class (\ref{mps:system1}) covers the classical linear equations (elliptic, parabolic and hyperbolic) and well-known nonlinear models (sin-Gordon, Burgers, Liouville,  Hopf, hyperbolic modification for the Burgers equation, equations with blow-up solutions).

For future handling the variety of models and their solutions, we need a convenient method for their identification. We propose the following descriptor for such objects:
 \begin{equation}\label{mps:descr}
 \begin{split}
EQ\bigl(&\tau,\alpha,\beta,\mu,\nu,\gamma,\xi,\theta,\chi; \\
\{&NONLINEAR\_FUNCTION: r,s,\varphi,k,\psi,h,b,f,I\};\\
\{&TITLE\_OF\_EQUATIONS\};\{TYPES\_OF\_SOLUTIONS\}\bigr)
\end{split}
\end{equation}

For instance, the descriptor for QHMB can be chosen in the  form
 \begin{equation*}
 \begin{split}
EQ\bigl(&1, 1, 1, 1, 1, 0, 0, 1, 0; \\
\{&NONLINEAR\_FUNCTION: \varphi, k,\psi, f \};\\
\{&QHMB\};\{compactons;blow-up;oscillations\}\bigr).
\end{split}
\end{equation*}
When the system of equations is considered, the coefficients and functions in (\ref{mps:system1}) should be assumed as  matrices. In particular, rewrite system (\ref{model}) as follows
\begin{equation*}
\begin{split}
 \frac{{\partial ^2 u}}{{\partial t^2 }} &=\rho^{-1} \left(e_1+3 e_3\left(\frac{\partial u}{\partial x}\right)^2\right) \frac{\partial
^2 u }{\partial x^2} + m \left(\omega^2(w-u)+\delta(w-u)^3\right), \\
 \frac{{\partial ^2 w }}{{\partial t^2
}} &=-\omega^2(w-u) - \delta(w-u)^3.
\end{split}
\end{equation*}
Then  its descriptor   is
 \begin{equation*}
 \begin{split}
EQ\Bigl(&\left(\begin{array}{cc}
1&0\\0&1\end{array}\right),
\hat 0,\hat 0, \left(\begin{array}{cc}
\rho^{-1}&0\\0&1\end{array}\right),\hat 0 ,\hat 0,\hat 0,\left(\begin{array}{cc}
m&0\\0&-1\end{array}\right) ,\hat 0; \\
\{&NONLINEAR\_FUNCTION: \\ &k=\left(\begin{array}{c}
e_1+3 e_3\left(\frac{\partial u}{\partial x}\right)^2 \\ 0\end{array}\right), f=\left(\begin{array}{c}
\omega^2(w-u)+\delta(w-u)^3 \\ \omega^2(w-u)+\delta(w-u)^3 \end{array}\right)\};\\
\{&EXAMPLE\_ONE\};\{solitary\, waves; compactones\}\Bigr).
\end{split}
\end{equation*}

Remark that the construction of descriptor helps us not only to classify the equations but at the statement of  problems. In particular, the Hopf equation can be considered as a limit case of hyperbolic Burgers equation.  

Let us also outline the types of solutions of such models. For generality, it is worth to mention that  certain equations from class (\ref{mps:system1}) possess interesting solutions, namely, autowave solutions to the Burgers equation,  singular solutions to the  Liouville equation, blow-up solutions to parabolic and hyperbolic equations with nonlinear sources, solitons for the sin-Gordon equation. Among the solutions to the  hyperbolic modification of the Burgers equation we encounter the packets of oscillations, compactons, autowaves.     

Another type of solutions is related to the multivaluedness. This means that the solutions have  several values in   given spatial point at fixed moment of time \cite{Mak_2015_Austria}. In particular, the Hopf equation  admits the solutions when their profiles  ``overturn"  and becomes multivalued. Similar behavior of wave solutions is observed in the Vakhnenko equation which admits the loop solutions. Studies of such solutions cause the introduction of new class of solutions named the ``foldons''. Foldons are the multivalued   autowaves. It is important that both the Hopf equation and Vakhnenko equation belong to class (\ref{mps:system1}) and their  multivalued solutions have analytical expressions. 

Up today we have hardly any methods to treat multivalued models. In this case the group analysis methods seem to be useful. To apply group methods, the multivalued differential equations should be considered as multivalued surfaces (geometric objects). The promising approach of multivalued model studies is concerned with the asymptotic transition from the singlevalued to multivalued models.  We encounter this  when transform the hyperbolic Burgers equation to the Hopf equation. Another way to  treat the multivaluedness is the expanding the inverse scattering transform.

It is worth to mention the development of novel approach dealing with the specific type of  inverse problems. The examples considered in the paper are concerned with analyzing the solutions  and their dependence on the parameters when the equations subjected to the initial and boundary data are available. Traditionally, such problems have been solved via the trial and error methods to find the self-similar solutions. But the inverse statement of the problem is possible. Indeed, we can take the functions $\tilde u (x,t)$ in such a way that they satisfy equation (\ref{mps:system1}), when the components in  (\ref{mps:system1})  are chosen correctly. 
In general case we would like to have the method when the set of functions $\tilde u_1(x,t)$, $\tilde u_2(x,t)$, ... , $\tilde u_n(x,t)$ satisfies  equation  (\ref{mps:system1}) subjected to proper set of initial conditions. This problem is similar to approaches in the theory of neural networks. The initial and boundary conditions are regarded as inputs of neural networks, whereas the functions $\tilde u_1(x,t)$, $\tilde u_2(x,t)$, ... , $\tilde u_n(x,t)$ considered as potential solutions are identified as outputs of neural networks. It is obvious that for arbitrary equation with fixed initial conditions the solutions    $ u_1(x,t)$, $u_2(x,t)$, ... , $ u_n(x,t)$ differ from the  ``desired'' set of function $\tilde u_1(x,t)$, $\tilde u_2(x,t)$, ... , $\tilde u_n(x,t)$. Then one can  find the values of equation's coefficients providing the small deviation $\Delta=\|\tilde u_i(x,t)-u_i (x,t)\|$. To realize this procedure, the iteration process of approaching to the equation's coefficients can be applied. This process is an analog of studying process in neural networks. We can use multivalued neural networks in these problems as well.

\section{Conclusion}

Summarizing, we presented the specific solutions occurred in quasilinear models for the carrying processes in nature. The results obtained were encouraged us to develop the systematic approach to the studies of nonlinear dynamical models within the framework of quasilinear second order equations and their solutions. We outlined  new ways to analyze quasi-linear models including the multivalued cases.

\end{document}